\documentclass[twocolumn,showpacs,preprintnumbers,prl]{revtex4}
\usepackage{graphics}
\usepackage{psfig,color}
\newcommand{\dd}{{\rm d}}
\begin{document}
\title{Prediction of anomalous diffusion and algebraic relaxations for long-range interacting systems, using classical statistical mechanics}
\author{Freddy Bouchet, Thierry
Dauxois\thanks{Thierry.Dauxois@ens-lyon.fr}}
\affiliation{Laboratoire de Physique, UMR CNRS 5672, {\'E}cole
Normale Sup{\'e}rieurs de Lyon, 46, all{\'e}e d'Italie, 69007
Lyon, France}

\date{\today}

\begin{abstract}
We explain the ubiquity and extremely slow evolution of non
gaussian out-of-equilibrium distributions for the Hamiltonian
Mean-Field model, by means of traditional kinetic theory. Deriving
the Fokker-Planck equation for a test particle, one also
unambiguously explains and predicts striking slow algebraic relaxation of the
momenta autocorrelation, previously found in numerical
simulations. Finally, angular anomalous diffusion are predicted
for a large class of initial distributions. Non Extensive Statistical Mechanics is shown to be  unnecessary for the interpretation of these phenomena.
\end{abstract}


\maketitle

Modern non-equilibrium statistical mechanics consider the study of
kinetic behavior of large ensembles of interacting particles, in
an attempt to relate macroscopic properties of a system to the
microscopic dynamics of its constituent. Recently there has been an increasing
interest on long-range interacting systems~\cite{Springer}. The
first reason is the broad spectrum of applications:
self-gravitating and Coulomb systems, vortices in two-dimensional
fluid mechanics~\cite{Chavanis}, wave-particles interaction and trapped charged
particles~\cite{Escande_Elskens}. Second, unexpected non gaussian
distributions~\cite{lrt2001} and non exponential relaxations for
autocorrelations~\cite{lrt2001,Pluchino,yamaPRE} have recently
been observed in such systems. This was a great surprise, indeed,
at variance with short-range interacting systems, as all the
particles contribute to the local field, mean-field predictions
are usually extremely good and often exact for the corresponding
equilibrium statistical mechanics. In accordance with this simple
picture, usual expectations were that these systems should exhibit
gaussian distribution functions and normal exponential
relaxations.

The  recent discovery of non gaussian distributions~\cite{lrt2001}
led to an intense and productive debate on the applicability~\cite{grigolini} of
usual Boltzmann-Gibbs statistical mechanics to long-range
interacting systems (see Tsallis {\em et al}~in
Ref.~\onlinecite{Springer}). Such non Gaussian distributions have
been fitted using Tsallis' distributions~\cite{lrt2001}. The
striking asymptotic algebraic behaviors for momentum
autocorrelations have also been fitted using q-exponential
functions~\cite{lrt2001,Pluchino}, derived from non extensive
statistical mechanics.

In this paper, revisiting this question, we
present for the first time analytical results explaining these
numerical findings, {\em without} using concepts of non-extensive
statistical mechanics.
In a recent paper~\cite{yama}, we explained
the ubiquitous occurrence of non equilibrium distributions by linking
them with stable Vlasov solutions. Here, we complete this
argument by explaining that the time scale for the evolution of
such non Gaussian distributions has to be larger than the number of particles~$N$. In an other recent paper, we have derived the Fokker Planck equation describing a single particle in an equilibrium bath \cite{freddy}. Here we generalize this result to out of equilibrium bath distributions. More importantly, we also report analytical explicit predictions for the  momentum
autocorrelations at large time, explaining its algebraic relaxation. Finally,
anomalous diffusion phenomena
in the asymptotic time limit were found in these systems and also
highly debated~\cite{lrr1999,yamaPRE}. This concept, raised in a
great variety of different fields such as geophysics, chemical
engineering and disordered media, was analytically understood in
only very few many-degrees-of-freedom systems, even if the
question of L{\'e}vy flights is very popular nowadays. Starting from the
microscopic dynamics of a long-range interacting system, we present in this letter, analytical predictions showing that there exists a full class of distributions which leads to anomalous diffusion. We give an explicit prediction of the anomalous diffusion exponent in terms of the bath distribution function. We inssit that such an exact prediction of this exponent, from the microscopic dynamics of a Hamiltonian N particle system is not common at all.

In this letter, we consider these questions for
the Hamiltonian Mean-Field model~\cite{antoni-95}
\begin{equation}
  \label{eq:hamiltonian}
  H_N = \frac{1}{2} \sum_{j=1}^{N} p_{j}^{2}  - \frac{1}{2N}
  \sum_{i,j=1}^{N} \cos(\theta_{i}-\theta_{j}),
\end{equation} since it
is nowadays thought to be the simplest model to study dynamical
and thermodynamic properties of system with long-range
interactions~\cite{Springer}. In addition to its pedagogical
properties, it corresponds to a simplification of one-dimensional
gravitational interactions and is an excellent first step before
the Colson-Bonifacio's model for free-electron Lasers~\cite{fel}.
Note that the  factor $1/N$ is the appropriate and classical {\em
mean field scaling}, relevant for long-range interacting
systems~\cite{spohn,BraunHepp}: indeed, the physically interesting
limit for such systems amounts to let the number of particles go
to infinity at fixed volume, by contrast with the usual
thermodynamic limit.

Let us consider the kinetics of the HMF model using the Klimontovich approach.
We have checked that an asymptotical expansion of the BBGKY hierarchy
leads to the same results.
The state of the $N$-particles system can thus be described by the {\em
discrete} single particle time-dependent density function
$ f_d\left(t,\theta,p\right)=
\frac{1}{N} \sum_{j=1}^N\delta
\left(\theta -\theta _{j}\left( t\right)\right)\delta \left(
  p-p_{j}\left( t\right) \right),$
where $\delta$ is the Dirac function, $(\theta,p)$ the Eulerian
coordinates of the phase space and  $(\theta_i,p_i)$ the
Lagrangian coordinates of the particles.
The dynamics is thus described by
the Klimontovich's equation~\cite{nicholson}
\begin{equation}
\frac{\partial f_d}{\partial t}+p\frac{\partial f_d}{\partial
\theta} -\frac{\dd V}{\dd \theta} \frac{\partial f_d}{\partial
p}=0 , \label{equationfdiscrete}
\end{equation}
where the potential $V$ that affects all particles is
$V(t,\theta)\equiv-\int_0^{2\pi}\!\!\dd
  \alpha\int_{-\infty}^{+\infty}\!\!\dd p\
  \cos(\theta-\alpha)\, f_d(t,\alpha,p) $.
This description of the Hamiltonian dynamics derived from
(\ref{eq:hamiltonian}) is {\em exact}: as the distribution is a sum of Dirac functions it contains the information on the position and velocity of all  particles. It is however too precise for usual physical quantities of interest but will be a key starting point for the derivation of
approximate equations, valid in the limit $N$ large and describing average quantities.

When $N$ is large, it is natural to approximate the discrete
density $f_d$ by a continuous one $f\left(t,\theta,p\right)$.
Considering an ensemble of microscopic initial conditions close to
the same initial macroscopic state, one defines the statistical
average $\langle f_d\rangle=f_0(\theta,p)$, whereas fluctuations
of probabilistic properties are of order $1/\sqrt{N}$. We will
assume that $f_0$ is any stable stationary solution of the Vlasov
equation. The discrete time-dependent density
function can thus be rewritten as
$f_d(t,\theta,p)=f_0(\theta,p)+\delta
f(t,\theta,p)/\sqrt{N}$,
where the fluctuation $\delta f$ is of zero average.
 We define similarly the averaged potential $\langle V\rangle$
and its corresponding fluctuations
$    \delta V(t,\theta)$
so that $V(t,\theta)=\langle V\rangle+\delta
V(t,\theta)/{\sqrt{N}}$. Inserting both expressions in
Klimontovich's equation~(\ref{equationfdiscrete}) and taking the
average, one obtains
\begin{eqnarray}
\frac{\partial f_0}{\partial t}+p\frac{\partial f_0}{\partial
\theta} -\frac{\dd \langle V\rangle}{\dd \theta} \frac{\partial
f_0}{\partial p}
&=&\frac{1}{N}\left\langle\frac{\dd \delta V
}{\dd\theta}\frac{\partial \delta f}{\partial p}\right\rangle.
\label{equationpourfzero}
\end{eqnarray}
The lhs is the Vlasov equation. The exact kinetic equation~(\ref{equationpourfzero}) suggests that for times  much smaller
than $N$, and stationary stable solutions $f_0$ for the Vlasov
equation, the rhs
term, corresponding to the fluctuation of the mean-field potential, can be neglected. This is confirmed, for any finite time, by the Braun and Hepp theorem  ~\cite{BraunHepp,spohn_livre}. 
The innumerable Vlasov stable stationary states,
 far from equilibrium~\cite{yama} (see~\cite{konishi-kaneko} for a first stability analysis), then explains the generic occurrence of
out-of-equilibrium distributions. These quasistationary states do not
evolve on time scales much smaller than $N$, explaining the
extremely slow relaxation of the system toward the statistical
equilibrium.

Let us now concentrate on stable homogenous distributions
$f_0(p)$, which are stationary since $\langle V\rangle=0$.
Subtracting Eq.~(\ref{equationpourfzero}) from
Eq.~(\ref{equationfdiscrete}) and using $f_d=f_0+\delta f/\sqrt{N}$, one gets
{\small
\begin{eqnarray}
\frac{\partial \delta f}{\partial t}+p\frac{\partial \delta
f}{\partial \theta}&-& \frac{\dd \delta V}{\dd \theta}
\frac{\partial f_0}{\partial p}
=
\frac{1}{\sqrt{N}}\biggl[\frac{\dd \delta V
}{\dd\theta}\frac{\partial \delta f}{\partial p}-\left\langle\frac{\dd \delta V }{\dd\theta}\frac{\partial \delta
f}{\partial p}\right\rangle \biggr].
\nonumber
\end{eqnarray}} 
For times much shorter than $\sqrt{N}$,  we
may drop the rhs encompassing quadratic terms in the fluctuations.
The fluctuating part $\delta f$ are then described, by the
linearized Vlasov equation (this is an other result of the Braun and Hepp theorem ~\cite{BraunHepp,spohn_livre}). This suggest to introduce the spatio-temporal
Fourier-Laplace transform
 of $\delta f$ and $\delta V$.
This leads to
\begin{eqnarray}\label{fouriertransformofdeltaVsuite}
\widetilde{\delta V}(\omega,k)
=-\frac{\pi \left(\delta_{k,1}+\delta_{k,-1}\right)}{\,\varepsilon(\omega,k)}\int_{-\infty}^{+\infty}\!\!\dd
p\ \frac{\widetilde{\delta f}(0,k,p)}{i(pk-\omega)} ,
\end{eqnarray}
where
 \begin{equation}
\varepsilon(\omega,k)=1+\pi{k}\left(\delta_{k,1}+\delta_{k,-1}\right)
\int_{-\infty}^{+\infty}\!\!\dd p\ \frac{\displaystyle\frac{\partial
f_0}{\partial p} }{(pk-\omega)}
\end{equation} is the dielectric permittivity.
 The evolution of the potential autocorrelation, can therefore be
determined
.  For homogeneous states, by symetry,
$\langle\widetilde{\delta V}(\omega_1,k_1)\widetilde{\delta
V}(\omega_2,k_2)\rangle=0$
 except if $k_1=-k_2=\pm1$.
One gets, after a transitory exponential decay, the general result
\begin{equation}
\left\langle  {\delta V}(t_1,\pm1){\delta
V}(t_2,\mp1)\right\rangle =\frac{\pi }{ 2}\int_{\cal C}\dd \omega
\ e^{-i\omega (t_1- t_2)} \ \frac{f_0(\omega)}{
\left|\varepsilon(\omega,1)\right|^2}. \label{correldeltaVfinal}
\end{equation}
This is an exact result, no approximation has yet been done.

A similar, although longer, calculation shows that the rhs of
Eq.~(\ref{equationpourfzero}) identically vanishes at order
$1/N$. This proves that Vlasov stable distribution function will
not evolve on time scales smaller or equal to $N$,
in agreement with the $N^{1.7}$ scaling law
which was numerically reported~\cite{Correlation,yama}.  This is the
first important result: {\it generic out of equilibrium
distributions evolve  on time scales much larger than
$N$}.

\begin{figure}
\resizebox{6truecm}{!}{\includegraphics{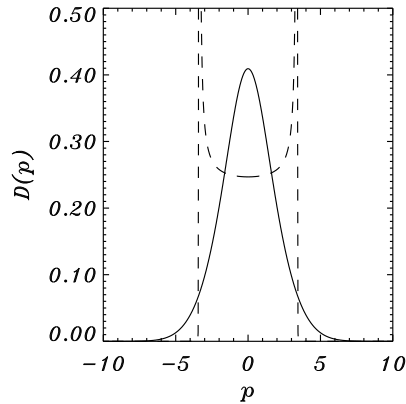}}
\caption{Diffusion coefficient $D(p)$ in the case $H_N/N=2$ for
a Boltzmann thermal bath (solid line) and a waterbag distribution
(dashed line). } \label{figplotDdep}
\end{figure}

Let us now consider relaxation properties of a test-particle, indexed
by 1, surrounded by a background system of $(N-1)$ particles with a
homogeneous distribution. The fluctuation of the potential is thus
\begin{equation}\label{fluctpotentialbis}
    \delta V(t,\theta)\equiv-
    \int_0^{2\pi}\!\!\!\!\!\!\!\dd
  \alpha\!\!\int_{-\infty}^{+\infty}\!\!\!\!\!\!\!\!\dd p\
  \cos(\theta-\alpha)\, \delta
f(t,\alpha,p) -\frac{1}{\sqrt{N}}\cos\left(\theta-\theta_1\right).
\end{equation}
Using the equations of motion of the test particle
and omitting the index 1 for the sake of simplicity, one obtains
$p(t)=p(0)-\int_0^{t}\!\!\dd u
{(\dd \delta V(u,\theta(u)))}/{(\dd \theta)}/{\sqrt{N}}$.
By introducing iteratively the expression of $\theta$ in
the rhs and expanding the derivative of the potential, one gets  the
result at order $1/N$. The key point is that this approach does not use
 the usual ballistic approximation. As a consequence, we obtain an exact result at order $1/N$. This is
 of paramount importance here to
treat accurately the {\em collective
effects}.
As the changes in the impulsion are small (of order
$1/\sqrt{N}$), the description of the impulsion stochastic process
by a Fokker-Planck equation is valid. This last equation is then
characterized by the time behavior of the first two moments
$\langle \left(p(t)-p(0)\right)^n\rangle$.
Using the generalization of
formula~(\ref{correldeltaVfinal}) when the effect of the test
particle is taken into account, one obtains in the large $t$-limit
\begin{eqnarray}\label{differentmoments1}
\langle
\left(p(t)-p(0)\right)\rangle&\stackrel{t\to+\infty}{\sim}&\!\!
\frac{t}{N}\left(\frac{\dd D}{\dd p}(p)
+\frac{1}{f_0}\frac{\partial f_0}{\partial p}D(p)\right)\quad\null\\
\langle
\left(p(t)-p(0)\right)^2\rangle&\stackrel{t\to+\infty}{\sim}&\!\!
\frac{2t}{N} \, D(p),\label{differentmoments2}
\end{eqnarray}
where the diffusion coefficient $D(p)$ can be written as
{\small
\begin{equation}
  D(p)
  =2 \,\mbox{Re}\int_{0}^{+\infty}\!\!\!\!\!\!\!\!\dd t\
  e^{ipt}\,\left\langle  {\delta V}(t,1){\delta
  V}(0,-1)\right\rangle
=   \ \frac{{{\pi^2}}f_0(p)}{
\left|\varepsilon(p,1)\right|^2}.\label{diffusioncoefficient}
\end{equation}
}
These results are the exact leading order terms in an expansion where $1/N$ is the small parameter. We obtained previously equivalent results, but restricted to equilibrium (gaussian) $f_0$~\cite{freddy}, by a rather different and instructive approach. In this last paper, a comparison of the diffusion coefficient $D(p)$ with N particle numerical simulations is presented, illustrating that both results are undistinguishable.  
Let us carry on by explicitly evaluating the diffusion coefficient
for a homogenous Gaussian distribution function
$f_{\rm g}(\theta,p)=\sqrt{{\beta}/{(2\pi)^3}} e^
{ -{\beta p^2}/{2}}$. In that case, after straightforward
calculations, one gets the expression derived in \cite{freddy}. The diffusion coefficient $D(p)$ has the asymptotic
expression $\sqrt{{\pi\beta}/{2}}\ e^{-{\beta p^2}/{2}}$ for large
$p$ values (see Fig.~\ref{figplotDdep}).

Interestingly, the method presented here, can be used for {\em any
Vlasov-stable out-of-equilibrium distributions}. For
instance, in Fig.~\ref{figplotDdep}, we present the result for the
waterbag distribution.

\begin{figure}
\resizebox{6truecm}{!}{\includegraphics{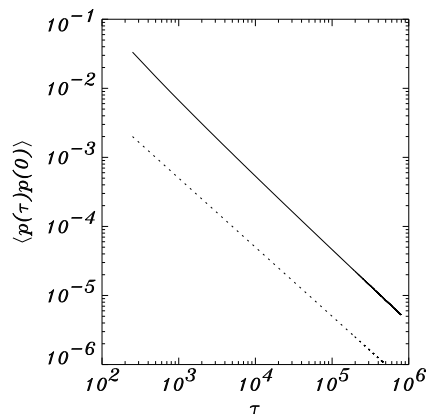}}
\caption{
The solid line represents the time evolution of the momentum autocorrelations obtained using the
numerical integration of the Fokker-Planck equation for a test
particle in a surrounding gaussian reservoir ($\beta=0.25$). Its slope
can hardly be distinguished from a  $1/ \tau$ law (dotted
line).
} \label{asympcorrel}
\end{figure}

Using time variable $\tau= t/N$ as suggested by
Eqs.~(\ref{differentmoments1}) and~(\ref{differentmoments2}), the
 Fokker-Planck equation describing the time evolution
of the distribution of the test particle is
\begin{equation}\label{fokkerplanckequation}
  \frac{\partial f_1(\tau,p)}{\partial \tau}=\frac{\partial }{\partial
  p}\left[D(p)\left(\frac{\partial f_1(\tau,p)}{\partial p}-\frac{1}{f_0}\frac{\partial f_0}{\partial
  p}f_1(\tau,p)\right)\right].
\end{equation}
We stress that this equation depends on the bath distribution $f_0$. It is valid both for equilibrium and out-of-equilibrium $f_0$, provided that $f_0$ is a stable stationary solution of the Vlasov equation. In the limit
$\tau\to\infty$ (more precisely $1 \ll \tau \ll N$),  the bracket vanishes: the pdf $f_1$ of the test
particle converges toward the quasi-stationary distribution $f_0$
of the surrounding bath. This is
in complete agreement with the result
that  $f_0$ is stationary for times scales of order~$N$.

For a large class of quasi-stationary distributions $f_0$, using
Eq.~(\ref{fokkerplanckequation}), we can compute the momenta
autocorrelation $\langle p(\tau)p(0)\rangle$,  fitted
numerically~\cite{Pluchino,Correlation,yamaPRE} with power laws,
stretched-exponentials or q-exponential. The second important
result of this letter is that {\it the time dependence of the
momentum autocorrelation function scale with $N$} as $\langle
p(t)p(0)\rangle = C(t/N)$ where $C(.)$ is a function.
  Let us first present the particular but
very important case of a test particle in contact with a {\em
gaussian distribution} $f_0$ (equilibrium bath).
Fig.~\ref{asympcorrel} shows the momenta autocorrelation $\langle
p(\tau)p(0)\rangle$, numerically computed from the Fokker-Planck
Eq.~(\ref{fokkerplanckequation}); it presents an unexpected  very slow
relaxation which can, numerically, hardly be distinguished from a
$1/\tau$-law. As shown below,  the very fast decrease of the
diffusion coefficient shown in Fig.~\ref{figplotDdep} is actually
the key point in these interesting and unusual properties of the
momenta autocorrelation.

One derive analytically the large time behavior of the autocorrelation function, for any Vlasov stable distribution $f_0$, using an asymptotic expansion inspired by~\cite{farago} (see also~\cite{micciche}). By introducing the appropriate
change of variable $x=x(p)$, defined by $\dd x/\dd
p=1/\sqrt{D(p)}$, one ends up with the constant diffusion
coefficient Fokker-Planck equation
\begin{equation}\label{fokkerplanckequationrescaledss}
  \frac{\partial \widehat{f}_1(\tau ,x)}{\partial \tau}=\frac{\partial }{\partial
  x}\left(\frac{\partial \widehat{f}_1(\tau ,x)}{\partial x}+\frac{\partial \psi}{\partial x}
  \widehat{f}_1(\tau ,x)\right),
\end{equation}
where $\psi(x) = - \ln\left(\sqrt{D(p)}f_0(p)\right)$.
Using $\varepsilon(p,1) \stackrel{|p|\to\infty}{\sim} 1
$, Eq.~(\ref{diffusioncoefficient}) shows that
$D(p) \stackrel{|p|\to\infty}{\sim} \pi^2 f_0(p)$.

Let us first consider distribution functions so that $f_0(p)
\stackrel{|p|\to\infty}{\sim} C\exp(-\gamma p^\delta)$, which
includes not only the gaussian  $(\delta = 2)$ and exponential
tails $(\delta = 1)$, but also stretched-exponential ones with
$\delta > 0$. Asymptotic analysis leads to $p(x)
\stackrel{|x|\to\infty}{\sim} 2/\gamma
\left(\ln(x)\right)^{1/\delta}$ whereas
\begin{equation}\label{aproxpsilargex}
\psi(x)\stackrel{x\to\pm\infty}{\sim}\alpha\ln |x|,\quad
\mbox{with} \ \alpha=3. \end{equation} Interestingly, this
log-potential is a limiting case between a pure diffusive process
leading to self-similar solutions and exponentially decreasing
$\widehat{f}_1$-solutions in a strongly confining potential.
  Asymptotic
 behavior (\ref{aproxpsilargex}) is still valid for distribution functions
 with algebraic tails, i.e. $f_0(p) \stackrel{|p|\to\infty}{\sim} Cp^{-\nu}$ where
 $\nu > 3$. However,   $\alpha=3\nu/(2+\nu)$ and  $p(x) \stackrel{|x|\to\infty}{\sim} C'x^{2/(2+\nu)}$.

By considering the ansatz
$\widehat{f}_1(x,t)=\varphi_\lambda(x)e^{-\lambda \tau}$, one gets
that  the spectrum of the Fokker-Planck operator corresponds to
only one bound state $\widehat{f}_1^0(x)=f_0$, normalizable for
$\alpha> 1$, with an associated eigenvalue located at the bottom
of the continuum: the absence of gap forbids {\em a priori} any exponential
relaxation.

As we are interested in the asymptotic large-$\tau$ limit, we
will restrict the analysis to the small-$\lambda$ regime. A matched
asymptotic procedure in the two distinct regions, $|x|>\ell$ and
$|x|<\ell$, solves the equation. In the first domain, introducing
$z=\sqrt{\lambda}\,x$ and $g_\lambda(z)=z\varphi_\lambda$, and
using the asymptotic result~(\ref{aproxpsilargex}), one ends up
with $z^2g_\lambda^{''}+ g_\lambda'(\alpha-2)
z+g_\lambda(2-2\alpha+z^2)=0$.
The solutions can be expressed in terms of Bessel functions of
order $\nu$, $J_\nu$ and $Y_\nu$, as
$g_\lambda(z)=A_{\lambda,\ell}\,
z^{{(3-\alpha)}/{2}}J_{{(\alpha+1)}/{2}}(z)+B_{\lambda,\ell}\,
z^{{(3-\alpha)}/{2}}Y_{{(\alpha+1)}/{2}}(z)
$. In the domain $|x|<\ell$, where one neglects the term
proportional to the vanishing eigenvalue $\lambda$,
the solution is
 $ \varphi_\lambda(x)=D_{\lambda,\ell}\,e^{-\psi(x)}+C_{\lambda,\ell}\,e^{-\psi(x)}\int_0^x\dd
  u \,e^{\psi(x)}
$.
In order to compute the momenta autocorrelation function, noting
that $p(x)$ is an odd function of $x$, we focus on odd
eigenstates, obtained by considering $D_{\lambda,\ell}=0$. By
taking care of the matching condition in $x=\ell$ and of the
normalization condition, one ends up with the scaling:
$A_{\lambda,\ell}\stackrel{\lambda\to0}{\sim}{\sqrt{\lambda}}/{2}$,
$B_{\lambda,\ell}\stackrel{\lambda\to0}{\sim} {{\lambda}}/{4}$ and
$C_{\lambda,\ell}\stackrel{\lambda\to0}{\sim}
C(l)\lambda^{(5/2)}$.

All these  results are finally useful to derive the
momentum auto-correlation. Indeed
using as initial condition
 $ \widehat{f}_1(x,0)\equiv N(\ell)p(x)e^{-\psi(x)}=\int_0^{+\infty}\dd
  \lambda\,\mu(\lambda)\varphi_\lambda(x)
$
 where $N(l)$ is the ground state normalisation factor and
$\mu(\lambda)=N(\ell)\int
\dd x\,p(x)
\varphi_\lambda(x) $, one gets
\begin{equation}\label{correlpp}
  \langle p(\tau)p(0)\rangle
  =\frac{1}{N(\ell)}\int_0^{+\infty}\dd
  \lambda\,\mu(\lambda)^2e^{-\lambda \tau}.
\end{equation}
Its limiting behavior in the $\tau\to\infty$ limit will be given
by the  behavior of $\mu(\lambda)$ when $\lambda\to0$, which is
itself determined by the large $|x|$ behavior of $p(x)$.

Let us be more specific in several important cases. The choice $p(x) \propto
(\ln x)^{1/\delta}$ which cor\-responds to distribution functions with
gaussian, exponential or stretched-exponential tails, leads to the
important result
\begin{eqnarray}
\langle p(\tau)p(0)\rangle
\stackrel{\tau\to+\infty}{\propto} \frac{(\ln \tau)^{2/\delta}}{\tau},
\label{correlinpasymp}
\end{eqnarray}
independent on $\alpha$. This very slow algebraic relaxation, with
logarithmic corrections, agrees with numerical simulation
shown in Fig.~\ref{asympcorrel}, for the Gaussian case $\delta=2$.

An extension of this approach to distribution function $f_0(p)$
with algebraic tails is of  prime interest. Following the same
steps in the case $p(x)=x^\eta$ ($\eta < (\alpha - 1)/2$) leads to
$\mu(\lambda)
\stackrel{\lambda\to0}{\propto}\lambda^{{(\alpha-2\eta-3)}/{4}}$
and to the limiting behavior
$\langle x^\eta(\tau)x^\eta(0)\rangle 
\stackrel{\tau\to+\infty}{\propto}
\tau^{\eta-{(\alpha-1)}/{2}}$.
Applying this result to the  algebraic decay of the distribution
function $f_0(p) \stackrel{|p|\to\infty}{\sim} Cp^{-\nu}$, with
$\nu > 2$ and using the corresponding asymptotic expression for
$p(x)$, discussed above, we get $\eta = 2/(2+\nu)$ and
$\alpha=3\nu/(2+\nu)$. We thus obtain the {\em algebraic decay}
\begin{eqnarray}
\langle p(\tau)p(0)\rangle \stackrel{\tau\to+\infty}{\propto}
\tau^{-\frac{\nu-3}{2+\nu}}. \label{correlinpasymp2}
\end{eqnarray}
This is the third important result:  the equilibrium distribution and
{\it generic out of equilibrium distributions lead to algebraic
large time behaviors of the momentum autocorrelation functions}.

 Finally, from the momenta correlations,  one
usually derives the angle diffusion $ \langle
(\theta(\tau)-\theta(0))^2\rangle=2D_\theta \,\tau$
where $D_\theta$ is defined via Kubo formula
$D_\theta=\int_0^{+\infty}\dd \tau\ \langle p(\tau)p(0)\rangle.$
However, asymptotic result~(\ref{correlinpasymp}) shows that
this integral diverges leading to small anomalous diffusion for
stretched exponential bath distributions. By contrast, for
distributions with algebraic tails,  since the exponent
$({\nu-3})/({2+\nu})$ in Eq.~(\ref{correlinpasymp2}) is smaller
than one, we predict {\em strong anomalous diffusion}. This is the
fourth important result.

In summary, this work confirms and explains the unexpected kinetic behavior of
long-range interacting systems: ubiquity of non gaussian
distribution and algebraic behavior for momenta autocorrelation
functions. Using traditional kinetic theory, we also predict strong
anomalous diffusion for angles for a large class of initial
distributions. Finally, let us anticipate
 that several physical systems with long-range interactions
should exhibit similar features. One would in particular quote
dynamics of vortices~\cite{chavanis} and cold atoms physics~\cite{lutz}.

\acknowledgments
We thank J. Barr{\'e}, P-H. Chavanis, I. Kourakis, S. Miccich{\`e}, S. Ruffo and C. Villani for helpful discussions.

\end{document}